\documentclass[a4paper,preprint,showpacs,superscriptaddress]{revtex4}
\usepackage{epsfig}
\begin{document}
\title{Long sandwich modules for photon veto detectors}
\author{N.~Yershov}\affiliation{Institute for Nuclear Research RAS, 117312 Moscow, Russia}
\author{M.~Khabibullin}\affiliation{Institute for Nuclear Research RAS, 117312 Moscow, Russia}
\author{Yu.~Kudenko}\affiliation{Institute for Nuclear Research RAS, 117312 Moscow, Russia}
\author{L.~Littenberg}\affiliation{Brookhaven National Laboratory, Upton, NY 11973, USA}
\author{V.~Mayatski}\affiliation{AO Uniplast, 600016 Vladimir, Russia}
\author{O.~Mineev}\affiliation{Institute for Nuclear Research RAS, 117312 Moscow, Russia}
\begin{abstract}
Long lead--scintillator sandwich modules developed
for the BNL experiment KOPIO are described. The individual  4~m long module
consists of 15 layers of 7~mm thick extruded scintillator  and 15 layers
of 1~mm lead absorber. Readout is implemented via WLS fibers glued
into grooves in a scintillator with 7~mm spacing and viewed from both ends
by the phototubes. Time resolution of 300~ps for cosmic MIPs was obtained.
Light output stability monitored for 2 years shows no degradation beyond
the measurement errors. A 4~m long C--bent sandwich module was also manufactured
and tested.
\end{abstract}
\maketitle
\section{Introduction}
\begin{figure}[hb] 
\begin{center} 
\epsfig{file=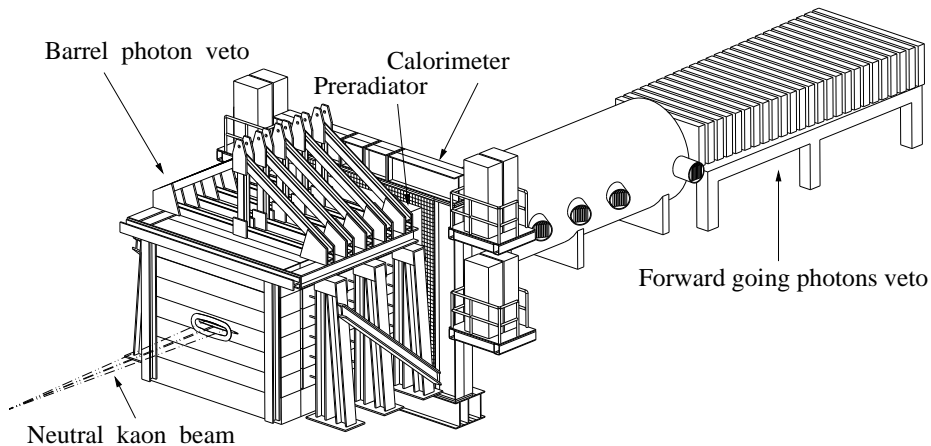,width=13.cm}
\end{center}
\caption{View of the KOPIO setup.}
\label{fig:layout926} 
\end{figure}

BNL experiment KOPIO \cite{e926} to study the rare kaon decay
$K^0_L\rightarrow\pi^0\nu\bar{\nu}$ represents a significant
experimental challenge.
The most difficult mode to suppress is $K^0_L\rightarrow\pi^0\pi^0$ which
can simulate $K^0_L\rightarrow\pi^0\nu\bar{\nu}$ if two of four photons are undetected.
To suppress the dominant background
from kaon decays a single photon veto capability must be up to an order
of magnitude better than measured in previous experiments for low
energy photons (below 50~MeV). High photon absorption efficiency is
required to achieve a detection efficiency of better than 0.9998 for 100~MeV photons.
High light yield over the entire detector volume is crucial to deal with
photonuclear
reactions that are likely to limit the achievable efficiency.
Since the momentum of the $K^0_L$ in KOPIO is determined
using a pulsed beam and time-of-flight method, the veto system must have good  time
resolution of about 200~ps for photon energies of 100--200~MeV while covering
the full solid angle around the kaon decay region. The cost of the veto system is
also a very important issue given the detector size of roughly $3\times3\times4$~m$^3$.
The scheme of the proposed experiment is shown in Fig.~\ref{fig:layout926}.

Lead--plastic scintillator sandwich veto counters are
being considered for the barrel veto detector which must have a thickness of
17~$X_{0}$ (radiation lengths). To read out light from the long scintillators, a two--ended
wavelength--shifting  fiber  readout technique will be used.
The two-ended readout provides good timing as well as redundancy for
failed channels. Similar detectors were used in the BNL experiment E787~\cite{e787}
(without WLS fibers) and the KEK experiment E391~\cite{e391}.
The extrusion method allows us to produce relatively cheap long
grooved scintillators which are well matched to the WLS fiber readout.
\section{Sandwich module design}
The results of tests of single extruded counters were reported in Ref.~\cite{nim1,nim2}.
The main results for extruded counters with WLS fiber readout are summarized
in Table~\ref{table:counters}. 
\begin{table}[htb]
\caption{ Parameters of  extruded polystyrene counters with 4.3~m long WLS fiber readout.
Fibers: multi--clad BCF99-29AA and single--clad BCF92 of 1~mm diameter.
 A counter made of BC404 scintillator is also shown for comparison.}
\begin{center}
 \begin{tabular}{ccccc}
 \hline
 Counter thickness  & Spacing  & Fiber type   & Light yield   & $\sigma_{t}$ \\
       mm           &    mm    &              &  p.e./MIP     &     ns          \\
 \hline
7  &   19                      & multi--clad &    11.2       &                 \\
7  &   10                      & multi--clad &    19.6       &       0.85      \\
7  &   10                      & single--clad&    14.4       &       0.87      \\
7  &   7                       & multi--clad &    26.2       &       0.71      \\
7  &   7                       & single--clad&    20.8       &       0.76      \\
3  &   10                      & multi--clad &     8.5       &       0.92      \\
\hline
7 (BC404)& 7              & multi--clad &    32         &       0.65      \\
\hline
 \end{tabular}
 \end{center}
 \label{table:counters}
\end{table}

Although the light attenuation length of extruded
polystyrene scintillator is about 30~cm, the one with WLS fiber readout
produces 0.8 of the light yield of BC404 scintillator. Single--clad (s.c.) and
multi--clad (m.c.) Bicron fibers provide practically the same time resolution.
Instead of using a wrapping material for a reflector we applied a novel technique:
the scintillator is etched by a chemical agent  that results in the formation of
a micropore deposit over the plastic surface, following which the diffuse film is
fixed in a settling tank. The deposit thickness (30-100~$\mu$m) depends on the
etching time. An advantage of this approach over the commonly used white diffuse
papers is the almost ideal contact of the reflector with the scintillator.
Moreover, it provides the option of gluing a lead sheet to the plastic covered by the chemical reflector
which facilitates assembling a sandwich unit. We tested a small size
sandwich assembly of 5 lead-plastic layers
glued together with a high viscosity polyurethane glue. After gluing
we found a light output reduction of about 6\%, but then no subsequent
degradation in the light yield
was observed for two months. To test the mechanical properties, a 4~m long dumb module (no fibers)
was assembled. Ten layers of 7~mm thick scintillator and 1~mm thick lead were glued
together.
An additional layer of 0.1~mm stainless steel foil was also glued to the first
plastic layer. The module sag under its own weight was measured to be 4~cm
after two weeks of testing. This value is close to the expected sag if the module
were a solid polystyrene.

Two straight modules have been manufactured at the Uniplast factory (Vladimir, Russia).
Extruded scintillator slabs 7~mm thick and  1~mm lead plates were fixed
together in a monolithic
block by an elastic polyurethane glue. The glue does not soak into the micropore chemical reflector.
The number of lead--plastic layers in a single module is 15. The module width is 150~mm.
Single-clad Bicron BCF-92 fibers are glued into 1.5~mm deep grooves which run along
the slab with 7~mm spacing with Bicron BC-600 optical glue.
The sandwich module is 4~m long. The WLS fibers which extend beyond the body of
the module are 4.5~m long.
FEU-115M phototubes with a green-extended photocathode view a bundle of 315
WLS fibers
at each end through silicone cookies, providing a two-end readout. The modules are wrapped
in black light isolation paper.
\section{Light output}
\begin{figure}[htb] 
\begin{center} 
\epsfig{file=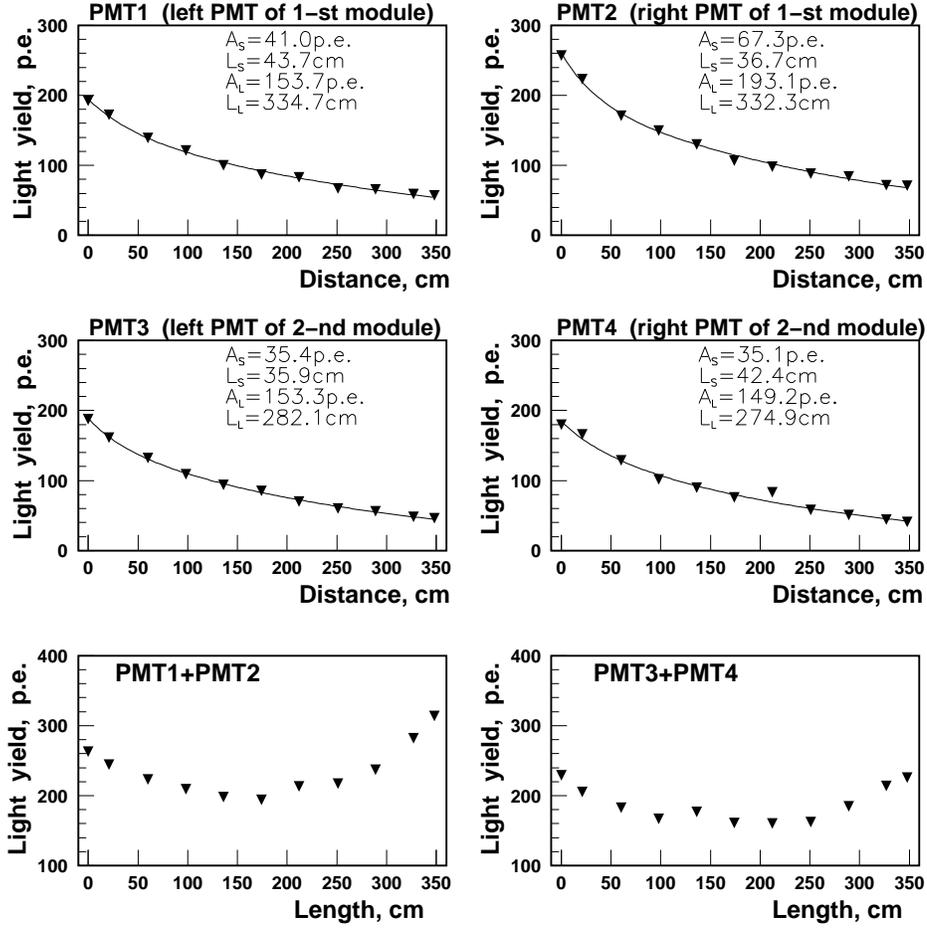,width=14.cm}
\end{center}
\caption{Light yield along the sandwich modules. The outermost points were taken at a distance of 25 cm from the module edges. The attenuation curves were fitted with $A=A_s\;exp(-\frac{x}{L_s})+A_L\;exp(-\frac{x}{L_L})$.}
\label{fig:adc-attenuation} 
\end{figure}

Two trigger counters selected the cosmic MIPs going through the sandwich modules
placed one above another. The upper trigger counter was 15~mm wide and 80~mm  long to localize
the hit position. The 4 PMT signals from the modules are sent to analogue fan-outs where they are distributed
to charge-sensitive ADCs and leading edge discriminators.
Before measurements of the module response,
single photoelectron peak of each phototube was calibrated
and these values are used to calculate the light yields. Previous experience
with FEU-115M PMTs shows that the single photoelectron
peak can drift by $\sim\pm$5\%. The accuracy of the light yield
measurements is determined mainly by this systematic error.
The light yield (l.y.) was scanned along the modules with a step of 30~cm. Results are shown in Fig.~\ref{fig:adc-attenuation}. 
The first module yielded about~195 p.e./MIP at the center,
corresponding to 9~p.e./MeV, and over 300~p.e./MIP near the ends.
The second module yielded 160~p.e./MIP at the center. The smaller l.y. is explained by the
different quality of available WLS fibers that resulted in a lower attenuation length
for the second module. The l.y. attenuation curves were fitted with a sum of two exponents. At
distances greater than 1~m the attenuation length of fast Bicron fibers was found to be 333~cm in the
first module, and 280~cm in the second module.
\begin{figure}[htb] 
\begin{center} 
\epsfig{file=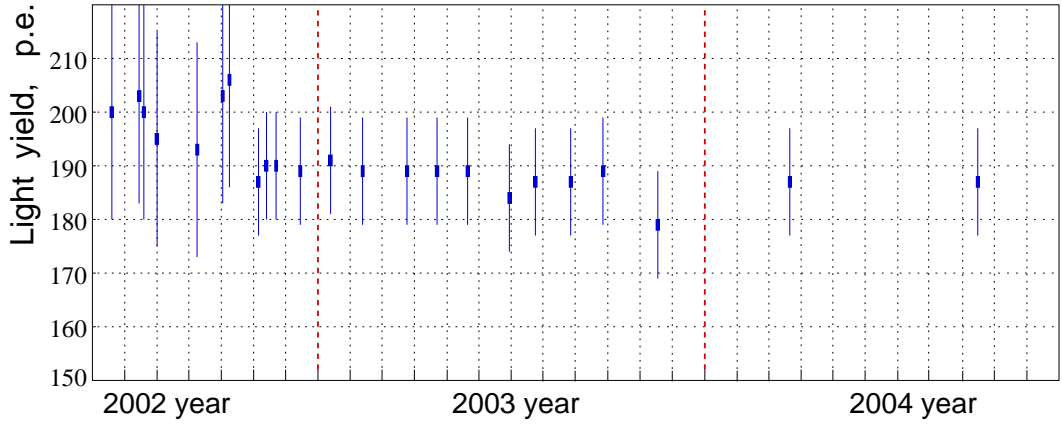,width=14.cm}
\end{center}
\caption{The light yield of the first sandwich module over a period of years.
The errors are determined by systematics.}
\label{fig:stability} 
\end{figure}
\begin{figure}[hbt]
\begin{center} 
\epsfig{file=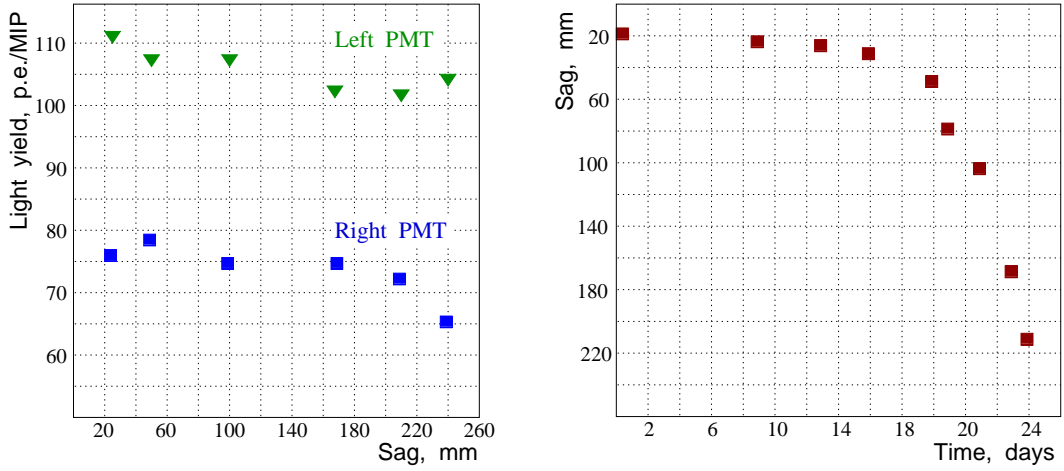,width=14.cm}
\end{center}
\caption{Light yield for each end of the second module versus its sag (left plot). Speed of module bowing (right plot).}
\label{fig:progib-combi} 
\end{figure}
The light output stability of the first module is shown in Fig.~\ref{fig:stability}. The reference point is the module center.  During the first 5 months of testing the trigger counters were moved along the module.
After that the test bench was fixed to periodically readout the data. As it can be seen in Fig.~\ref{fig:stability} the l.y. is stable over 2 years within the 5\% systematic error. The second module was subjected to the sagging test. During test the both ends rested upon the supports,
and the sag was measured as deviation of the module center from the straight horizontal level. Results are shown in Fig.~\ref{fig:progib-combi}. 
The initial deviation was 2~cm. After 2 weeks the
flexure under its own weight led to fast uncontrolled bowing. However the sag effect on the light yield is rather weak.
\section{Timing}
\begin{figure}[htb] 
\begin{center} 
\epsfig{file=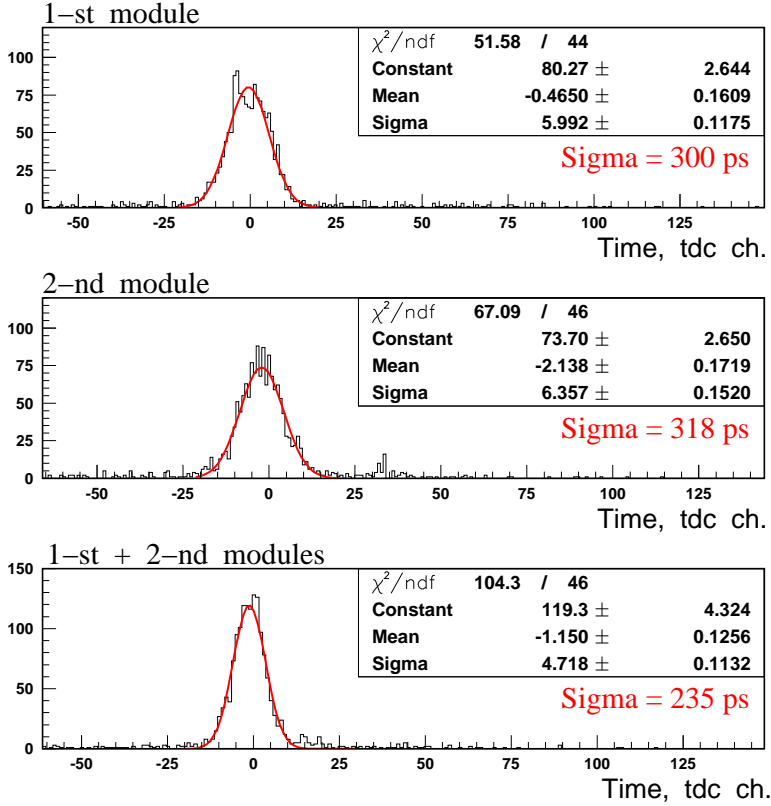,width=12.cm}
\end{center}
\caption{Time resolutions obtained at the centers of sandwich modules.
The combination $(TDC_{left}-TDC_{right})/2$ was
used to make the timing spectra for the modules. Half-sum of these spectra produced a time
resolution of 235~ps. The TDC scale is 50 ps/ch.}
\label{fig:timing} 
\end{figure}

\begin{figure}[hbt] 
\begin{center} 
\epsfig{file=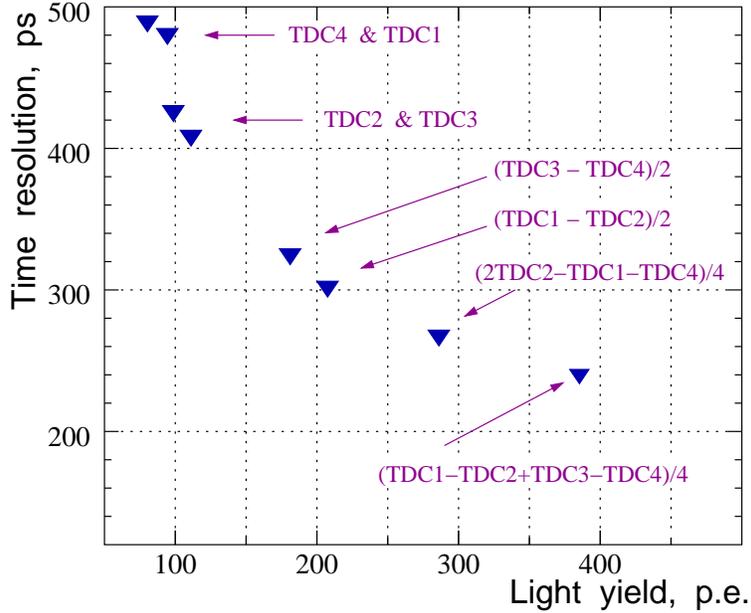,width=10.cm}
\end{center}
\caption{Time resolution (rms) vs light output. TDC1 and TDC2 are the ends of first module,
TDC3 and TDC4 are the ends of second module.}
\label{fig:tdc-attenuation} 
\end{figure}

A time-amplitude correction was applied to all signals from the sandwich modules.
To suppress the timing spread caused by the trigger counters we used
the combination $(TDC_{left}-TDC_{right})/2$. The timing spectra obtained in
this way at the module centers are shown in Fig.~\ref{fig:timing}.
The time resolution is 300~ps for the first module, and
slightly worse for the second module. Cosmic MIPs deposit 21~MeV in the
scintillator of a single module.
The combination $(TDC1_{left}+TDC2_{left}-TDC1_{right}-TDC2_{right})/4$ produces
a resolution of 235~ps.
The resolution dependence on the light output is demonstrated in Fig.~\ref{fig:tdc-attenuation}.
\begin{figure}[htb] 
\begin{center} 
\epsfig{file=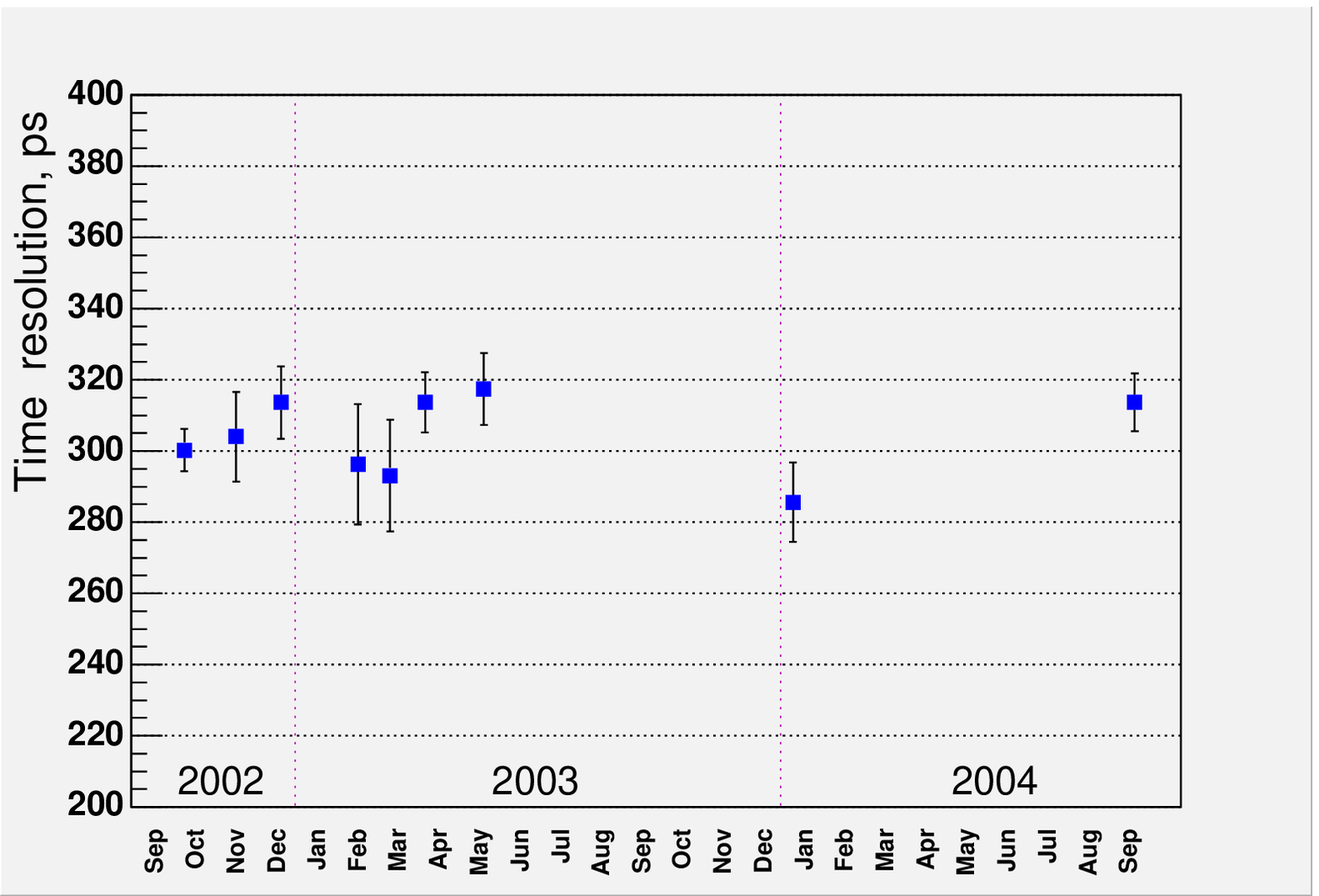,width=12.cm}
\end{center}
\caption{Time resolution (rms), measured at the center of the first module over years. The points
fluctuate around 300~ps within the time fitting error.}
\label{fig:tdc-stability} 
\end{figure}
Only central values are shown as the error is determined by unspecified systematic factors.
Fitting the points with the root square law yields
$\sigma_t[ps]=4330/\sqrt{E[p.e.]}+2.7$, where E is the light yield in photoelectrons.
Taking into account that the visible fraction for a module is 0.4 the time resolution
for photons can be brought to $\sigma_t[ps]=72ps/\sqrt{E_{\gamma}[GeV]}+2.7$
\begin{figure}[htb] 
\begin{center} 
\epsfig{file=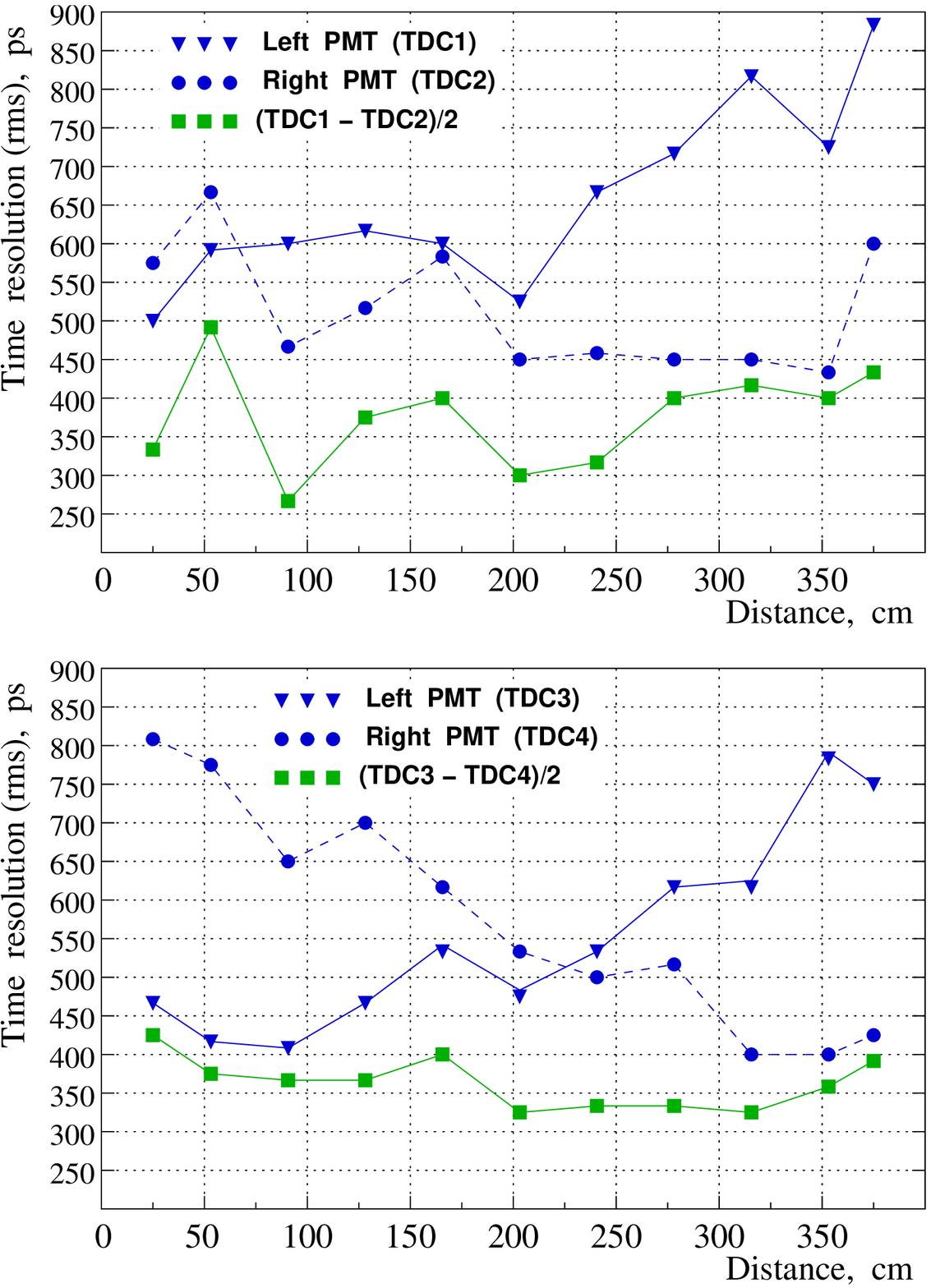,width=10.cm}
\end{center}
\caption{Time resolution (rms) along the module. Upper plots show results for the first module, lower plots for the second one. }
\label{fig:tdc-uniform12} 
\end{figure}
Fig.~\ref{fig:tdc-stability} shows the time
resolution for the first module over two years. The resolution over the module
length is shown in Fig.~\ref{fig:tdc-uniform12}. 
Applying the straight $(TDC_{left}-TDC_{right})/2$ combination the best timing is obtained at the module center. Near the ends the resolution have to be calculated with a more complicated algorithm.
\section{C-bent sandwich module}
\begin{figure}[htb] 
\begin{center} \epsfig{file=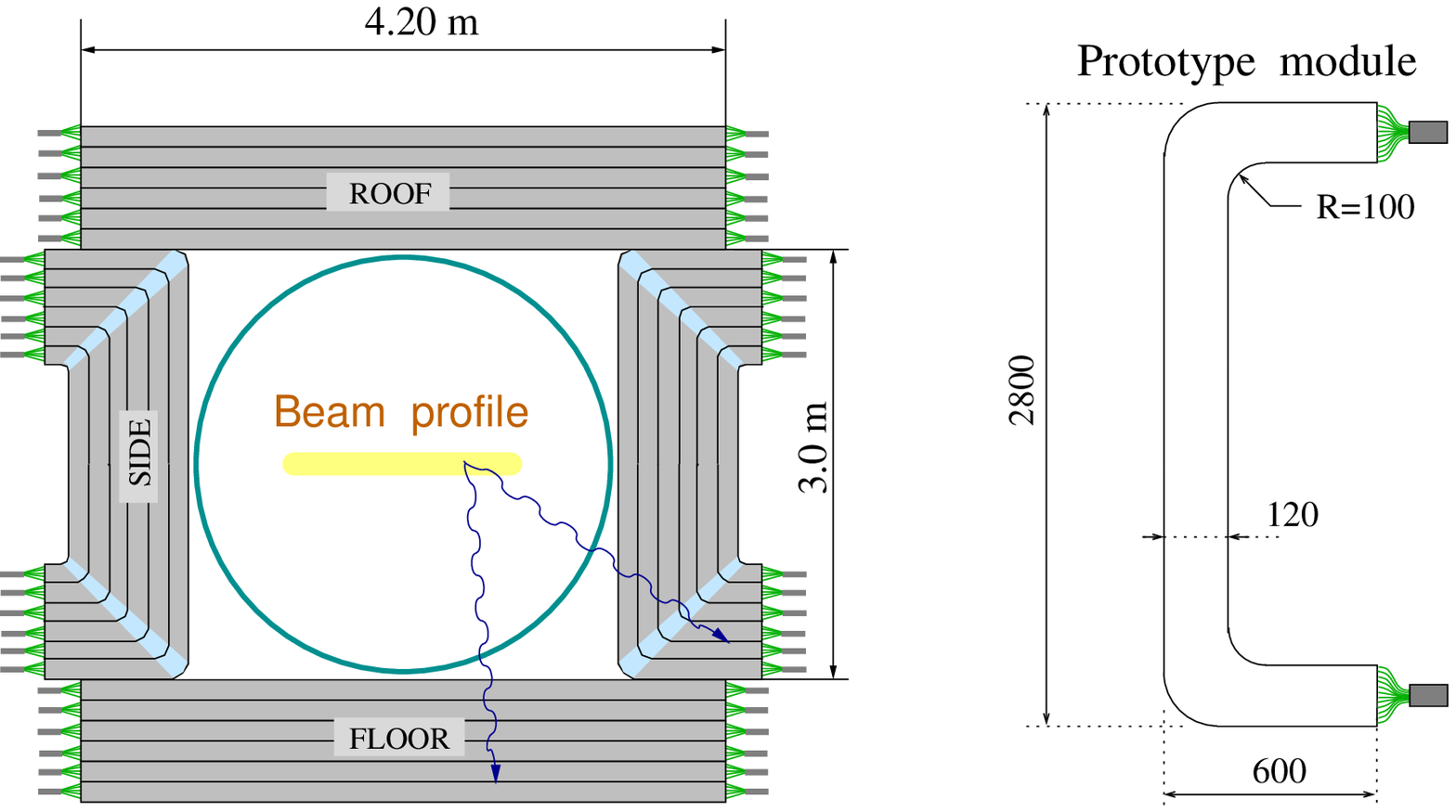,width=10.cm}
\end{center}
\caption{Schematic view of the photon barrel veto with C-bent sandwich modules as the side walls. Dimensions of the tested prototype C-module are shown on the right.}
\label{fig:side} 
\end{figure}

Full photon veto coverage was the motivation to design the C-bent sandwich modules which can
be used to build side walls in the photon veto barrel. The barrel roof and floor are made of
straight sandwich modules. Phototube readout for the barrel side walls
would create a dead space
degrading the photon detection efficiency. C-bent modules were proposed to resolve
the problem as shown in Fig.~\ref{fig:side}. 
There will be no longer be any projective gaps
for photons emerging from the flat kaon beam. The C-modules are installed
vertically, since for the
horizontal alternative the bent part of the modules produces directions with low radiation thickness.

To check the method a few scintillator slabs of 1~m length with WLS readout were produced.
Their light output was measured  and then the slabs were bent using the different thermal
treatment. The light output was measured again with the same phototube. The best
result shows no light yield drop, the average loss in l.y. for other bent slabs was about 5\%.

\begin{figure}[htb]
\begin{center} 
\epsfig{file=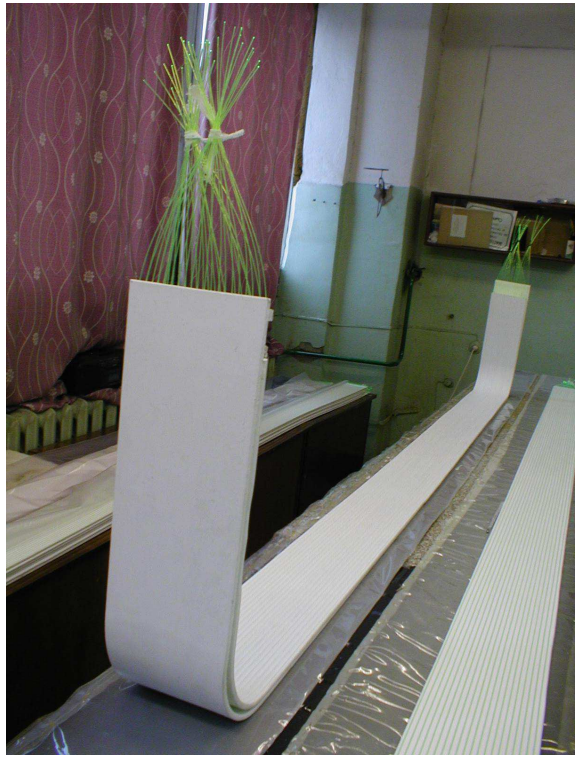,width=10.cm,angle=-90}
\end{center}
\caption{The scintillator slabs after bending. A straight slab before bending
is shown on the right.}
\label{fig:c-slab} 
\end{figure}

\begin{figure}[htb] 
\begin{center} 
\epsfig{file=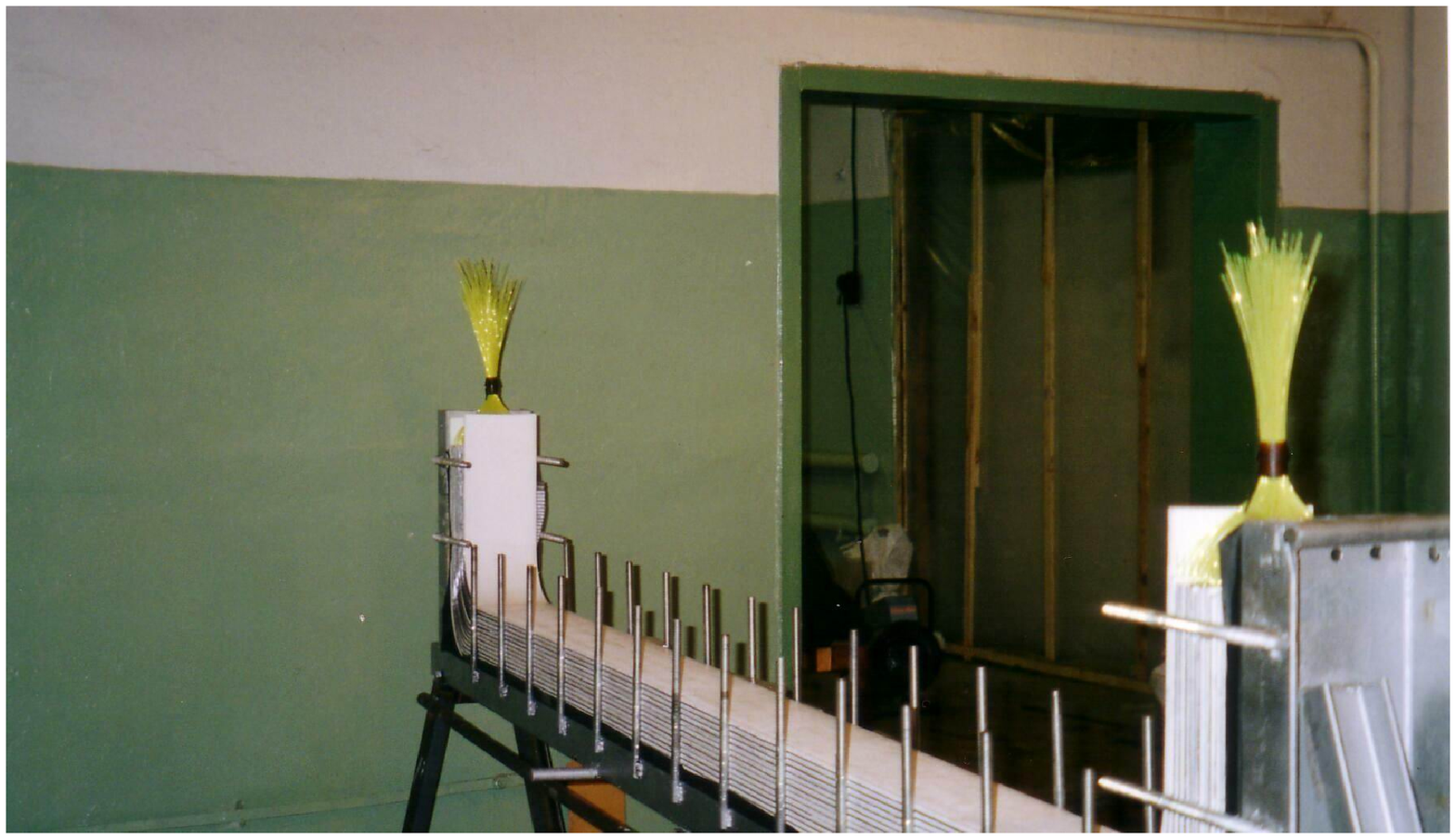,width=10.cm}
\end{center}
\caption{The assembly of 15 C-bent scintillator slabs interleaved with lead layers in a frame jig.}
\label{fig:c-jig} 
\end{figure}

\begin{figure}[htb]
\vspace{.5cm}
\begin{center} 
\epsfig{file=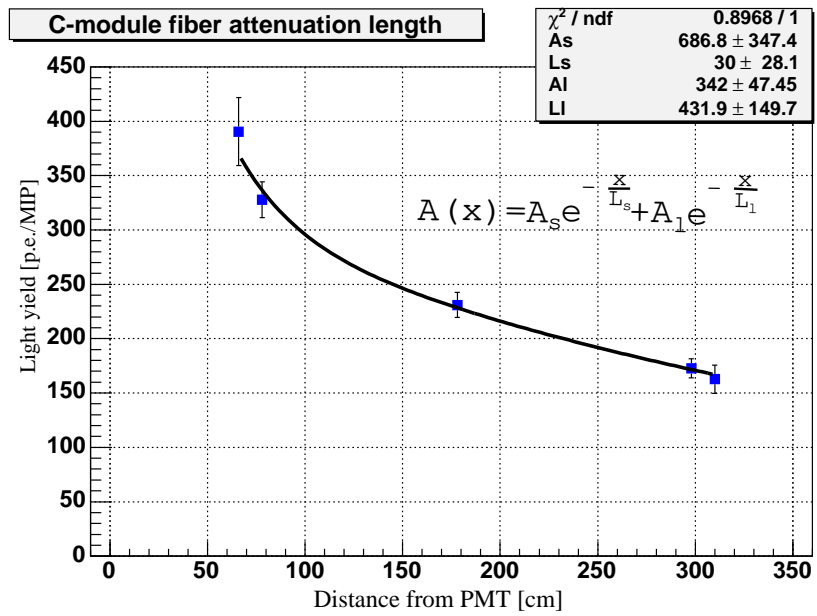,width=12.cm}
\end{center}
\caption{Light yield from cosmic MIPs in C-bent module from a single end. L$_S$ -- short
attenuation length, L$_L$ -- long attenuation length.}
\label{fig:c-attenuation} 
\end{figure}

\begin{figure}[htb] 
\begin{center} 
\epsfig{file=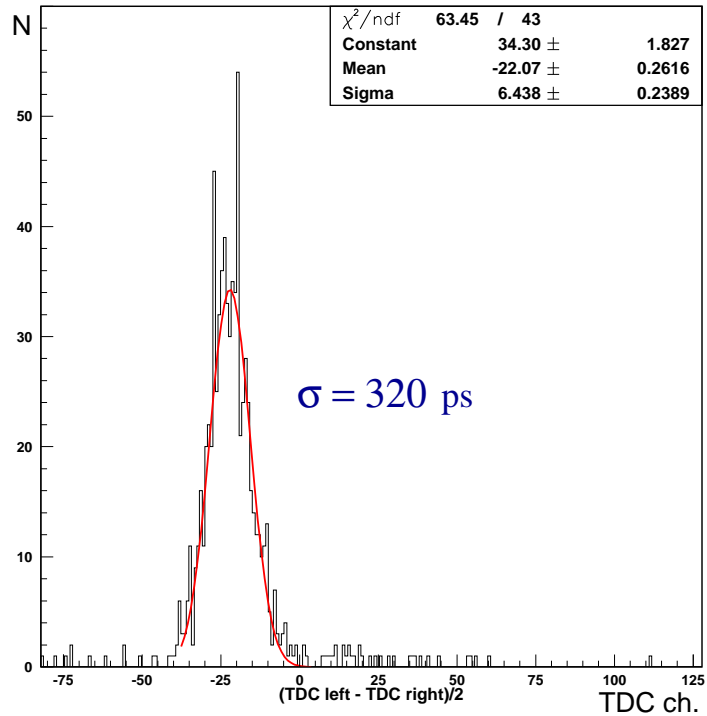,width=11.cm}
\end{center}
\caption{Time spectrum of the C-bent module with cosmic muons at the center after time-amplitude correction. A TDC channel is equal to 50 ps.}
\label{fig:c-time} 
\end{figure}

The prototype module was bent by the following procedure:
\begin{enumerate}
\item First, 7~mm thick slabs were produced with length from 3.3 to 4~m applying
the standard methods. Two-mm deep grooves run along the slab with spacing of 7~mm.
Y11 Kuraray WLS fibers of 1~mm diameter were glued into the straight grooved slabs
of extruded polystyrene scintillator covered by chemical reflector.
\item The bent slabs were prepared by
locally heating the bending place with a warm iron cylinder. The cylinder temperature
was stabilized using water flow through the cylinder. The thermostat controls this temperature.
\item Bending over the cylinder with 10~cm radius is done such that the fibers
run along the inner curvature
of the slab. Optimum conditions were selected to heat up and cool down the scintillator with
fibers avoiding its shape deformation. The bent slabs are shown in Fig.~\ref{fig:c-slab}.
\item The prototype module had been assembled in a frame jig from 15 bent scintillator slabs
and 15 layers of 1~mm thick lead by gluing them together. The module in a jig is shown in
Fig.~\ref{fig:c-jig}. 300 WLS fibers are squeezed then at both ends into the collet connectors.
\end{enumerate}
The prototype module was tested with cosmic rays in the central part. The light yield of the
C-module is 500~p.e./MIP from both ends or about 24~p.e./MeV. It is a factor of 2.5 larger than
the light output for the straight sandwich modules due to multi-clad Y11 fibers. The light output
from a single end along the straight part of the C-module is shown in Fig.~\ref{fig:c-attenuation}.
The long attenuation length is measured to be 4.3~m. This value is close to Kuraray data that
demonstrates the quality of fibers after thermal bending.
Although Y11 fibers have a longer decay time than Bicron ones, the obtained time resolution
of $\sigma_{t}$=320~ps (see Fig.~\ref{fig:c-time}) is close to the result (300~ps) for the straight
modules. After 8 months the measured light yield at the center show no change.
\section{Conclusion}
Long prototype photon veto detectors for the BNL experiment KOPIO were manufactured and
tested with cosmic rays.
The sandwich modules consist of 15~layers of 7~mm thick extruded scintillator
and 15~layers of 1~mm lead absorber. One of the modules was bent to have a C-shape.
Readout is implemented with WLS fibers. The average yield was measured with
cosmic rays to be about 9~p.e./MeV in the center for straight modules with fast single-clad
Bicron fibers.
The light yield for the C--bent module is 2.5 times larger due to multi-clad Y11 Kuraray fibers.
The time resolution for all modules is about 300~ps. The light output of one of the straight
modules was monitored over 2 years and shows no degradation beyond the measurement errors.
The light output of the thermally bent C-module is also stable after 8 months.

\section{Acknowledgment}
 We gratefully acknowledge NSF support through the KOPIO Experiment.

\end{document}